\documentstyle[12pt,epsfig]{article}
\def\Bll{$B \rightarrow X_s l^+ l^-$}

\def\BTT{B \rightarrow X_s \tau^+ \tau^-}
\def\phm{$\phi_\mu$}
\def\pha{$\phi_{A_0}$}
 
\begin{document}
\bigskip
{\large\bf
\centerline{Supersymmetric CP Violation in \Bll }
\centerline{ in Minimal Supergravity Model}
\bigskip
\normalsize

\centerline{Chao-Shang Huang,~~~Liao Wei}
\centerline{\sl Institute of Theoretical Physics, Academia Sinica,
      P.O.Box 2735,}
\centerline{\sl Beijing 100080,P.R.China}
\bigskip

\begin{abstract}
Direct CP asymmetries and the CP violating normal polarization of lepton in inclusive
decay \Bll are investigated
in minimal supergravity model with CP violating phases. The contributions
coming from exchanging neutral Higgs bosons are included.  
 It is shown that the direct CP violation in branching ratio, $A_{CP}^1$, is of  ${\cal{O}}(10^{-3})$
 for $l=e, \mu, \tau$. The  CP violating normal polarization for l=$\mu$ can reach 0.5
percent when tan$\beta$ is large (say, 36).
For $l=\tau$ and in the case of large $\tan\beta$, the direct CP violation in backward-forward asymmetry,
 $A_{CP}^2$, can reach one percent, the normal polarization of $\tau$ can be as large as 
a few percent, and both are sensitive to the two CP violating 
phases, \phm ~ and \pha, and consequently it could be possible 
to observe them (in particular, the normal polarization of $\tau$)  in the future B factories.
\end{abstract}

\newpage

\section{Introduction}
\label{sec:introduction}
CP violation has so far only been observed in K system. It is one of the
goals of the B factories presently under construction to discover and examine
CP violation in the B system. CP violation is originated from the CKM matrix
~\cite{ckm} in the standard model(SM) and new sources of CP violation may appear in
extensions of SM. In the constrained minimal supersymmetric standard model, i.e
the minimal supergravity model(mSUGRA), besides the standard CP violating phase $\delta_{CKM}$,
there are two new CP violating phases, which may be chosen as the phase of $\mu$ (\phm)
 and the phase of $ A_0$ (\pha), that can't be
rephased away when the universality of soft terms is assumed 
at unification scale\cite{dgh}.

It is well-known that the supersymmetric (SUSY) CP violating phases are constrained by
the experiments on the electric dipole moments (EDMs) of neutron and electron~\cite{dgh,edm}.
SUSY contributions to EDMs will exceed the current experimental limits on the EDMs of
neutron and electron unless either the SUSY phases are rather much smaller ($\leq 10^{-2}$)~
\cite{edm} or sfermion masses of the first and second generations are very large ($>$ 1 Tev)
~\cite{nat}. However, large sfermion masses may be incompatible with bounds on the relic 
density of a gaugino-type LSP neutrilino. Recently, a third possibility  to evade the EDM 
constraints has been pointed out~\cite{IN,bgk}. That is, various contributions to EDM cancel with 
each other in significant regions of the parameter space, which allows SUSY phases to be of order 
one and sparticles are relatively light. It is found that in the mSUGRA with small tan$\beta (\leq 3)$ , 
the phase $|\phi_{\mu}| \leq \frac{\pi}{10}$ while the phase $\phi_{A_0}$ remains essentially unconstrained, 
by combining cosmological and EDM constraints~\cite{FO}. Similar results have also been obtained in Refs.~\cite{aad}.
In this paper we shall investigate 
effects of SUSY phases on $B\rightarrow X_s l^+l^-$ (l=e, $\mu, \tau$) assuming the third 
possibility to evade the EDM constraints (i.e choosing the parameters in the region of the 
parameter space where cancellations happen). We extend the analyses of the EDM constraints to the large
tan$\beta$ ($\geq$ 20) case. It is found that the cancellations are insufficient to make the EDMs of electron 
and neutron satisfy the experimental limits if $\phi_{\mu}$ is of order one and the sparticle spectrum is 
below ${\cal{O}}$ (1 Tev). That is, in the large tan$\beta$ case  $\phi_{\mu}$ must be $\leq 10^{-2}$ in order to satisfy the 
experimental limits of EDMs of electron and neutron and have a relatively light sparticle spectrum.

Effects of SUSY CP violating phases on the branching ratio of $B\rightarrow X_s l^+l^-$ have
been examined~\cite{goto,gk}. In this paper we study direct CP asymmetries and the CP violating
normal polarization of lepton in $B\rightarrow X_s l^+l^-$ (l=e, $\mu, \tau$) in mSUGRA with
CP violating phases. The direct CP asymmetry of this mode in the SM is unobservably small.
Thus, an observation of CP violation
in this mode would signal the presence of physics beyond SM. \\

\section{N=1 supergravity and CP violation phases}
\label{sec:N=1}
In mSUGRA it is assumed that the soft SUSY breaking terms, which are originated from the 
gravitational interaction, are universal at the high energy (GUT or Planck) scale. So there
are only five free parameters at the high energy scale: $M_{1/2}$, the mass
of gauginos; $A_0$, the trilinear couplings; $B$, the bilinear couplings; $M_0$, the universal 
masses for all scalars, as well as $\mu$, the Higgs mass parameter in superpotential($\mu$ and $A_0$ are defined as in \cite{IN}).
In general $A_0$, $B$, $\mu$ and $M_{1/2}$ are complex. However, not all the phases are
physical. It is possible to rephase away the phase of
$M_{1/2}$ and to make $B \mu$ real by redefinition of the fields and
by R transformation~\cite{dgh}. So there are only two physically independent phases left
, which can be chosen as $\phi_{A_0}$ and $\phi_{\mu}$. The breakdown of 
electroweak symmetry via radiative effect allows one to determine the 
magnitude of $\mu$ and $B$ at electroweak scale. Therefore, one has four real parameters
($M_0, M_{1/2}, |A_0|, tan\beta$) and two phases ($\phi_{\mu}$, \pha) finally.

Mass spectra of sparticles, flavor mixing, and soft term parameters at the EW scale can be
determined by solving the renormalization group equations (RGEs) running from GUT scale to 
EW scale. In order to see the running of the phases we record  the equations
for trilinear terms and $\mu$ (we neglect the effects of $A_i$ of 1st and 2nd generation
because of their corresponding very small Yukawa couplings)~\cite{rge}:
\begin{eqnarray}
&& \frac{dA_u}{dt} = \frac{1}{4 \pi}( \frac{16}{3}\alpha_3 M_3+
3\alpha_2 M_2+\frac{13}{15}\alpha_1 M_1+3 Y^t A_t), \nonumber \\
&& \frac{dA_d}{dt} = \frac{1}{4 \pi}( \frac{16}{3}\alpha_3 M_3+
3\alpha_2 M_2+\frac{7}{15}\alpha_1 M_1+3 Y^b A_b+Y^{\tau}
 A_{\tau}) , \nonumber \\
&& \frac{dA_e}{dt}= \frac{1}{4 \pi}( 3\alpha_2 M_2+\frac{9}{5}
\alpha_1 M_1+ 3 Y^b A_b+ Y^{\tau} A_{\tau}),  \nonumber \\
&& \frac{dA_t}{dt} = \frac{1}{4 \pi}( \frac{16}{3}\alpha_3 M_3+
3\alpha_2 M_2+\frac{13}{15}\alpha_1 M_1+6 Y^t A_t+Y^b A_b), \nonumber \\
&& \frac{dA_b}{dt} = \frac{1}{4 \pi}( \frac{16}{3}\alpha_3 M_3+
3\alpha_2 M_2+\frac{7}{15}\alpha_1 M_1+Y^t A_t+6 Y^b A_b+Y^{\tau}
 A_{\tau}) , \nonumber \\
&& \frac{dA_{\tau}}{dt}= \frac{1}{4 \pi}( 3\alpha_2 M_2+\frac{9}{5}
\alpha_1 M_1+4 Y^{\tau} A_{\tau}+3 Y^b A_b),  \nonumber \\
&& \frac{d\mu}{dt} = \frac{\mu}{8 \pi}(-\frac{3}{5} \alpha_1 -3 \alpha_2
+Y^\tau +3 Y^b+3 Y^t)
\end{eqnarray}
where $\alpha_i=\frac{g_i^2}{4 \pi}$, $Y^i=\frac{y_i^2}{4 \pi}(i=t,b,\tau)$,
$g_i$ are the gauge coupling constants, $y_i$ are Yukawa couplings
and $t=ln(Q^2/M_{GUT}^2)$.
It is explicit from the equations that the phase of $\mu$ does not run and both the magnitudes 
and phases of $A_i$ evolve with $t$. \\

\section{Constraints on the parameter space from EDMs of electron and neutron and $B\rightarrow X_s\gamma$}
\label{sec:constraints}
The cancellation mechanism for suppression of the EDMs of electron and neutron in mSUGRA with SUSY phases
have been pointed out. In the small tan$\beta$ (say, tan$\beta$=3) case the region of the parameter space in which
the EDMs of electron and neutron satisfy the experimental limits \cite{data}
\begin{eqnarray}
|d_e|< 4.3\times 10^{-27} ecm,
\end{eqnarray}
and
\begin{eqnarray}
|d_n|< 6.3 \times 10^{-26} ecm
\end{eqnarray}
have been analyzed \cite{IN,bgk}. We make similar analyses and confirm their results. We extend the analyses to the large
tan$\beta$ ($\geq$ 20) case. It is found that the cancellations are insufficient to make the EDMs of electron and neutron
satisfy the experimental limits if $\phi_{\mu}$ is of order one and the sparticle spectrum is below ${\cal{O}}(1 Tev)$. Thus we have to
give up the phase $\phi_{\mu}$ of order one and search for  cancellations for $\phi_{\mu}\leq 10^{-2}$ in the large
tan$\beta$ case if we stick to the low sparticle spectrum. We scan the parameters $M_0, M_{1/2}, |A_0|$ and $\phi_{A_0}$
in the range of $300\leq M_0,M_{1/2}\leq 800, 100 \leq |A_0|\leq 1200$, and $0\leq\phi_{A_0}\leq 2\pi$ for fixed values of tan$\beta$ and $\phi_{\mu}$ in
the range of tan$\beta \geq 20$ and $\phi_{\mu} \leq 10^{-2}$.
 It is found that there are significant regions in which sparticle spectrum are below
${\cal{O}}$ ( 1Tev) and the EDM constraints are satisfied. The result for a set of typical values of  the parameters $M_0, M_{1/2}, |A_0|$  and 
$\phi_{\mu}$ ($M_0= 400, M_{1/2}= 550, |A_0|= 750, tan\beta= 36$ and $\phi_{\mu}=\pm \pi/1000$ ) is shown in fig.1, where EDME and EDMN represent the EDMs of
electron and neutron respectively. The bounds between the two horizontal lines
in the figures represent the experimental limits of the EDMs of electron and neutron. One can see from the figure that the 
electron EDM (fig.1a) imposes a constraint on $\phi_{A_0}$ and the neutron EDM (fig1b) imposes no constraints on $\phi_{A_0}$.
The experimental constraint on the electron EDM is more stringent than that on the neutron EDM when there exist
cancellations between different components, a conclusion similar to the small tan$\beta$ case. The EDMs of electron and neutron as functions of 
$\phi_{A_0}$ in the small tan$\beta$ case have also been given in fig.1. There are constraints on $\phi_{A_0}$, which is different from ref. \cite{FO}.
The main reason is that we use the new data of the EDMs of electron and neutron which are quite smaller than the old data they used.\\

Because we shall pay attention to the case of large tan$\beta$ we should also consider the contributions
arising from Barr-Zee mechanism~\cite{bz}. It is found that~\cite{ckp}
\begin{eqnarray}
(\frac {d_f}{e})_{BZ}= Q_f \frac {3 \alpha}{64 \pi^2}\frac {R_f m_f}{m_A^2} \sum_{q=t,b}
\xi_q Q_q^2 [ F(\frac {m_{\tilde{q}_1}^2}{m_A^2})- F(\frac {m_{\tilde{q}_2}^2}{m_A^2})],
\end{eqnarray}
where $R_f= cot\beta (tan\beta)$ for $I_{3f}$=1/2 (-1/2), and
\begin{eqnarray}
\xi_t=\frac{sin 2\theta_{\tilde{t}} m_t Im(\mu e^{i\delta_t})}{sin^2\beta v^2},
\xi_b=\frac{sin 2\theta_{\tilde{b}} m_b Im(A_b e^{i\delta_b})}{sin\beta cos\beta v^2}
\end{eqnarray}
with $\delta_f=arg(A_f+R_f \mu^{*})$, and F(x) can
be found in Ref.~\cite{ckp}, due to the two loop diagram contributions. The parameters chosen
in our numerical calculations of $B\rightarrow X_s l^+l^-$ satisfy the constraints from EDMs including the above two loop contributions. 

 It is well-known that $B\rightarrow X_s\gamma$ puts a stringent constraint on the parameter space of mSUGRA without SUSY
CP violating phases \cite{bg,hy}. With SUSY CP violating phases, we calculate $B_r(B\rightarrow X_s\gamma)$. The branching ratio
as function of $\phi_{A_0}$ is shown in fig.2, for the values of other SUSY parameters same as those in fig.1. As can be seen
from the fig.2, region in the range $0\leq \phi_{A_0} \leq 2\pi$ is allowed by the experimental limit of   $B_r(B\rightarrow X_s\gamma)$. 
Since under the choices given here results are almost the same for sign of
$\phi_{\mu}$ switched and other parameters unchanged, we only show the cases of
positive $\phi_{\mu}$ in fig.2.

\section{Formulas for \Bll}
\label{sec:formulas}
Neglecting the strange quark mass,the matrix element governing
the process \Bll is given as follows~\cite{dhh,hy}:
\begin{eqnarray}
M &=& \frac{G_F \alpha}{\sqrt{2} \pi} V_{tb}{V_{ts}}\hspace{-5pt}^* \bigg[ 
{C_8}^{eff} \overline{s}_L \gamma_{\mu} b_L \overline{\tau} \gamma^\mu
\tau +C_9(m_b) \overline{s}_L \gamma_\mu b_L \overline{\tau} \gamma^\mu 
\gamma_5 \tau-2 C_7(m_b) m_b \overline{s}_L i \sigma_{\mu \nu}
\frac{q^\nu}{q^2} b_R \overline{\tau} \gamma^{\mu} \tau \nonumber \\
&& +C_{Q1}(m_b) \overline{s}_L b_R \overline{\tau} \tau +
C_{Q2}(m_b) \overline{s}_L b_R \overline{\tau} \gamma_5 \tau \bigg]
\end{eqnarray}
where~\cite{gsw}
\begin{eqnarray}
&{C_8}\hspace{-2pt}^{eff} = C_8(m_b)+ (3 C_1(m_b)+ C_2(m_b)) \bigg[
g(\frac{m_c}{m_b},\hat{s})+ \lambda_u (g(\frac{m_c}{m_b},\hat{s})
-g(\frac{m_u}{m_b},\hat{s})) \nonumber \\
&+\frac{3}{\alpha^2} \kappa \sum_{V_i=\psi'}
\frac{\pi M_{V_i} \Gamma (V_i \rightarrow \tau^+ \tau^-)}
{M^2\hspace{-2pt}_{V_i}-q^2-i M_{V_i} \Gamma_{V_i}} \bigg],\\
&g(z,\hat{s})=-\frac{4}{9} lnz^2+ \frac{8}{27}+ \frac{16}{9} 
\frac{z^2}{\hat{s}}-\left\{ \begin{array}{ll}
 \frac{2}{9} \sqrt{1-\frac{4 z^2}{\hat{s}}} (2+\frac{4 z^2}{\hat{s}})
 \bigg[ ln ( \frac{1+\sqrt{1-4 z^2/ \hat{s}}} {1-\sqrt{1-4 z^2/ \hat{s}}})
 +i \pi \bigg], & \textrm{$4 z^2 < \hat{s}$} \\
 \frac{4}{9} \sqrt{\frac{4 z^2}{\hat{s}}-1} (2+\frac{4 z^2}{\hat{s}})
arctan \bigg( \frac{1}{\sqrt{4 z^2/ \hat{s} -1}} \bigg), 
 & \textrm{$4 z^2 > \hat{s}$}
\end{array} \right.
\end{eqnarray}
with $q=p_{l^+}+p_{l^-}$,$\hat{s}=q^2/{m^2}_b$ and $\lambda_u=\frac{V_{ub}
V_{us}^*}{V_{tb}V_{ts}^*}$. The final two terms in eq.(6) come
from exchanging neutral Higgs bosons (NHBs).

The QCD corrections to coefficients $C_i$ and $C_{Q_i}$ can be incorporated
in the standard way by using the renormalization group equations.
Since no NLO corrections to $C_{Q_i}$ have been given,we use the leading
order corrections to $C_i$ and $C_{Q_i}$ although the NLO corrections to
$C_i$ have been calculated.They are given as below \cite{dhh}:
\begin{eqnarray}
&& C_7(m_b) = \eta^{- \frac{16}{23}} C_7(m_W)+\frac{8}{3}(\eta^
{-\frac{14}{23}}-\eta^{-\frac{-16}{23}}) C_{8G}(m_W)+ C_{2}(m_W)\sum_{i=1}^8
 h_i \eta^{a_i}-0.012 \eta^{-\frac{16}{23}} C_{Q_3}(m_W) , \nonumber \\
&& C_8(m_b) = C_8(m_W)+\frac{4 \pi}{\alpha_s(m_W)} \big[ \frac{8}{87}
(1-\eta^{-\frac{29}{23}})- \frac{4}{33} (1-\eta^{-\frac{11}{23}})] C_2(m_W),
\nonumber \\
&& C_9(m_b) = C_9(m_W), \nonumber \\
&& C_{Q_i}(m_b) = \eta^{- \gamma_Q / \beta_0} C_{Q_i}(m_W), ~~i=1,2 ,
\nonumber \\
&& C_1(m_b) = \frac{1}{2} \big (\eta ^{- \frac{6}{23}} -\eta ^\frac{12}{23}
\big) C_2(m_W), \nonumber \\
&& C_2(m_b) = \frac{1}{2} \big(\eta ^{- \frac{6}{23}} +\eta ^\frac{12}{23}
\big) C_2(m_W)
\end{eqnarray}
where $\gamma_Q=-4$ is the anomalous dimension of $\overline{s}_L b_R$~\cite{h},
$\beta_0 = 11-2 n_f/3$, $\eta = \alpha_s(m_b)/ \alpha_s(m_W)$, $C_2(m_W)
=-1$, and $C_i(m_W)$ (i=7,8,9) and $C_{Q_i}(m_W)$ (i=1,2,3) can be found in
Refs.~\cite{bb} and ~\cite{hy}, respectively (since flavor changing contributions from 
gluino-downtype squark loop and neutrilino-downtype squark loop are very small compared 
to those from chargino-uptype squark loop
in mSUGRA~\cite{bb,hy}, we neglect them in this paper).\\

With the matrix element (eq.(6)), it is easy to derive the invariant dilepton mass distribution 
as follows~\cite{dhh}:
\begin{eqnarray}
&& \frac{{\rm d} \Gamma(\BTT)}{{\rm d} \hat{s}} = B(B \rightarrow X_c l 
\overline{\nu}) \frac{\alpha^2}{4 \pi^2 f(\frac{m_c}{m_b})} (1-\hat{s})^2
\Big( 1-\frac{4 t^2}{\hat{s}} \Big)^\frac{1}{2} \frac{ \big|V_{tb} V_{ts}
\hspace{-4pt}^* \big|^2}{\big| V_{cb} \big|^2} D(\hat{s}),\nonumber \\
&& D(\hat{s}) =  4 \big|C_7 \big|^2 (1+\frac{2}{\hat{s}}) (1+\frac
{2 t^2}{\hat{s}})+\big|{C_8}\hspace{-4pt}^{eff} \big|^2 (2 \hat{s}+1)
(1+\frac{2 t^2}{\hat{s}})+ \big|C_9 \big|^2 \big[ 1+2 \hat{s}
+ (1-4 \hat{s}) \frac{2 t^2}{\hat{s}} \big] \nonumber \\
&& +12 Re({C_8}\hspace{-4pt}^{eff} C_7\hspace{-4pt}^* ) (1+\frac{2 t^2}
{\hat{s}})+ \frac{3}{2} \big|C_{Q_1} \big|^2 (1-\frac{4 t^2}{\hat{s}})
\hat{s}+ \frac{3}{2} \big|C_{Q_2} \big|^2 \hat{s}+ 6 Re(C_9 C_{Q_2}
\hspace{-8pt}^* ) t
\end{eqnarray}
where $t=m_{\tau}/ m_b$,$B(B \rightarrow X_c l \overline{\nu})$ is the 
branching ratio, and f(x) is the phase space factor:$f(x)=1-8 x^2+8 x^6-x^8
-24 x^4 lnx$. Backward-forward assymmetry can also be calculated to be:
\begin{eqnarray}
&& A(\hat{s}) = \frac{\int^{1}_{0}dz \frac{d^2\Gamma}{d\hat{s} dz}-
\int^{0}_{-1}dz \frac{d^2 \Gamma}{d\hat{s} dz}}
{\int^{1}_{0}dz \frac{d^2\Gamma}{d\hat{s} dz}+\int^{0}_{-1}dz
\frac{d^2 \Gamma}{d\hat{s} dz}}
 = \frac{E(\hat{s})}{D(\hat{s})} , \nonumber \\
&& E(\hat{s}) = 3 \sqrt{ 1-\frac{4 t^2}{\hat{s}}}
Re( C_8\hspace{-5pt}^{eff} C_9\hspace{-2pt}^* \hat{s}
+2 C_7 C_9 \hspace{-2pt}^* +C_8\hspace{-5pt}^{eff} C_{Q_1}\hspace{-5pt}^* t
 + 2 C_7 C_{Q_1}\hspace{-5pt}^* t)
\end{eqnarray}

The direct CP assymmetries in decay rate and backward-forward
assymmetry for $B\rightarrow X_s l^+l^-$ and $\bar{B}\rightarrow \bar{X_s} l^+l^-$ are defined by
\begin{eqnarray}
&& A_{CP}\hspace{-12pt}^1 \hspace{12pt}(\hat{s}) = \frac{{\rm d} \Gamma 
/{\rm d} \hat{s} - {\rm d}\overline{\Gamma} /{\rm d} \hat{s}}{{\rm d}
\Gamma /{\rm d} \hat{s} +{\rm d} \overline{\Gamma}/{\rm d} \hat{s}} ,\\
&& A_{CP}\hspace{-12pt}^2 \hspace{12pt} (\hat{s}) = \frac{A(\hat{s})-
\overline{A}(\hat{s})}{A(\hat{s})+\overline{A}(\hat{s})}
\end{eqnarray}

In SM the direct CP violation can only arise from the interference of
non-trivial weak phases which are contained in CKM matrix elements.
Therefore, it is suppressed  by  the ratio of CKM matrix elements,
$\frac{V_{us}^*V_{ub}}{V_{ts}^*V_{tb}}\sim {\cal{O}}(10^{-2})$. The CP
assymmetry in the branching ratio is predicted to be of the order of 
$10^{-3}$ \cite{KSAH}, which is unobservably small. Thus, an observation of  CP 
violation in this mode would signal the presence of new physics. In mSUGRA 
without new CP violating phases, the alignment of masses of the first two
generation squarks causes a cancellation and only contributions proportional
to $\frac{m_c}{m_W}$ or $\frac{m_u}{m_W}$ (coming from couplings of chargino-
-right handed up type squarks-down type quarks) remain for the imaginary
parts of Wilson coefficients.  But as the mixings of left and right handed
squarks of 1st and 2nd generations are negligibly small, imaginary parts
of Wilson coefficients can not be large. So without new
CP violating phases, mSUGRA will induce CP violating effects in the same
order as those in SM in these processes. In mSUGRA with new CP violating phases,
one may expect larger CP asymmetries due to the presence of new phases of order one. 

Another CP violating observable in $B\rightarrow X_s l^+ l^-$ is the normal polarization of 
the lepton in the decay, $P_N$, which is the T-violating projection of the lepton spin onto
the normal of the decay plane, i.e $P_N \sim \vec{s}_l\cdot (\vec{p}_s\times \vec{p}_{l^-})$
~\cite{Lee}. A straightforward calculation leads to
\begin{eqnarray}
  P_N = \frac{3 \pi}{4} \sqrt{1-\frac{4 t^2}{\hat{s}}} \hat{s}^\frac{1}{2}
Im\Big[2 C_8 \hspace{-2pt}^{eff*} C_9 t +4 C_9 C_7\hspace{-2pt}^*
\frac{t}{\hat{s}}+ C_8\hspace{-2pt}^{eff*} C_{Q_1}+ 2 C_7\hspace{-2pt}
^* C_{Q_1}+ C_9\hspace{-2pt}^* C_{Q_2} \Big] \Big/ D(\hat{s}) 
\end{eqnarray}
$P_N$ have been given in the ref.\cite{11}, where they give only two
terms in the numerator of $P_N$.

We may find from eq.(12) that contributions to $P_N$ are of imaginary
parts of product of two Wilson coefficients, i.e the product of real
part of one Wilson coefficient and imaginary part of another one. So 
compared to $A_{CP}^1$, $P_N$ can be larger when the imaginary parts of
some relevant Wilson coefficients have significant values. Because the
normal polarization is proportional to the mass of lepton (we remind that
$C_{Q_i}$ (i=1,2) is proportional to $m_l$~\cite{hy}), it will be 
unobservably small for l=e. However, for 
l=$\mu, \tau$, when $C_{Q_i} (i=1,2)$ get significant values, $Im(C_8 \hspace
{-2pt}^{eff*} C_{Q_1}\hspace{2pt})$ and $Im(C_9 \hspace{-2pt}^* C_{Q_2}
\hspace{2pt})$, as well as $Im(C_8\hspace{-2pt}^{eff*} C_9)$, will be main
contributions, which can make $P_N$ large. \\

\section{numerical results}
\label{sec:numerical}
In the numerical calculations in mSUGRA the SUSY parameters are taken as
\begin{eqnarray}
M_0= 800 Gev, M_{1/2}= 180 Gev, |A_0|= 350 Gev, tan\beta=2, \phi_{\mu}=\pm \pi/30;  \nonumber
\end{eqnarray}
or
\begin{eqnarray}
M_0= 400 Gev, M_{1/2}= 550 Gev, |A_0|= 750 Gev, tan\beta=36, \phi_{\mu}=\pm \pi/1000 \nonumber
\end{eqnarray}
with $\phi_{A_0} \sim {\cal{O}}(1)$
which satisfy the constraints from EDMs of neutron and electron,
as well as the constraint from $b\rightarrow s \gamma$. We follow Ref.~\cite{hy}
for detailed procedures of calculation.\\

With these choices the direct CP violation in branching ratio is of the order $
{\cal{O}}(10^{-3})$ for l=e, $\mu$ and $\tau$, i.e the same order as that in SM,
 thus is hard to be observed. However, as pointed out in ref.~\cite{goto},
 Im$C_7$ can be as large as $C_7$ of SM at some values of the parameters and 
consequently a CP violation significantly larger than the result may probably
be obtained. The results for $A_{CP}^2$ at $\hat{s}= 0.76$ and $P_N$ which is experimentally 
measured, are shown in Figs. 3 and 4, where we do not show the regions of \pha in which the constraints from EDMs
of electron and neutron and $B_r(B\rightarrow X_s \gamma)$ are satisfied (the regions can be easily read from figs.1,2). 
One can see from fig.3 that in the large $tan\beta$ case the direct CP asymmetry in backward-forward asymmetry, $A_{CP}^2$,
for $l=\tau$ can be 0.5 to 1 percent in the most of range of \pha and sensitive to \pha, which is not easy to
 be observed. $A_{CP}^2$ for l=e, $\mu$ is as small as $A_{CP}^1$. This large difference between $\tau$ and e, $\mu$ is due to
that the contributions of exchanging NHBs are proportional to the
mass square of lepton (see eq.(11) ).

Fig. 4 shows the CP violating normal polarization. The polarization is almost equal to zero
for l=e no matter how tan$\beta$ is large or small because of negligible smallness of electron 
mass. For l=$\mu$, it can reach only 0.5 percent and is sensitive to \phm, \pha when tan$\beta$
is large. For l=$\tau$, it is close to 2 percent and not sensitive to \phm, \pha for small
tan$\beta$ and can be as large as 4 percent and sensitive to \phm, \pha for large tan$\beta$,
which could be observed in the future B factories with $10^8-10^{12}$ B hadrons per year~\cite{bs}. 
The reason is that $C_{Q_i}$ can be neglected for tan$\beta$=2 and $C_9$, $C_8^{eff}$ depend
on \phm, \pha weakly. But for tan$\beta$ = 36, $C_{Q_i}$'s have large imaginary parts and  
strongly depend on \phm, \pha. There is a cancelation between contributions
from $Im(C_8 \hspace{-2pt}^{eff*} C_{Q_1}\hspace{2pt})$ and $Im(C_8\hspace
{-2pt}^{eff*} C_9)$ in the large tan$\beta$ case, since when $\phi_{\mu}$ is
around 0 the real part of
$C_{Q_1}$ is of the opposite sign of the real part of $C_9$.
Combined with the suppression coming from the enhancement
of $D(\hat{s})$ induced by large $C_{Q_1}$ and $C_{Q_2}$, this cancellation
may cause even a lower value of $P_N$ (for some values of  \pha) than
that in SM, as can be seen from fig.4. The sensitivity of the normal polarization to \phm, \pha can be used to discriminate small tan$\beta$
from large tan$\beta$.\\
 
In summary, we have analyzed EDMs and $B\rightarrow X_s\gamma$ constraints on the parameter space in mSUGRA 
with CP violating phases, in particular, in the case of large tan$\beta$.
we have calculated the direct CP asymmetries and CP violating normal polarization
for the rare decays $B\rightarrow X_s l^+l^-$ (l=e, $\mu$, $\tau$) in the model.
When \phm ~and \pha ~are of ${\cal{O}}(0.1)$ and ${\cal{O}}(1)$ respectively, the direct CP asymmetries in branching ratio
 for l=e, $\mu, \tau$ are about $10^{-3}$, i.e the same order as that in SM.
So it is hopeless to observe $A_{CP}^1$ in mSUGRA even with  large CP 
violating phases no matter how tan$\beta$ is large or not. 
For l=e, $\mu$, $A_{CP}^2$ is as small as $A_{CP}^1$ .
  In the case of large tan$\beta$ the CP violating normal polarization of muon in $B\rightarrow X_s
\mu^+ \mu^-$ can reach 0.5 percent . For $B\rightarrow X_s \tau^+ \tau^-$, $A_{CP}^2$ can reach
one percent and the normal polarization of $\tau$  can be as large as 
a few percent when $\tan\beta$ is large (say, 36), and consequently it is possible to 
observe the normal polarization. Thus, a few percent CP asymmetry would be discovered in $B\rightarrow X_s \tau^+\tau^-$ if 
the nature gives us a CP violating SUSY with large (say, $\geq$ 30) tan$\beta$.  \\

Recently, it is shown that if gaugino masses at high energy scale are nonuniversal there exist two
additional phases which can make cancellations happened easier than the universal case \cite{bgk}.
It is found that the phases may be large while certain approximate relations hold among the mass
parameters and phases, resulting in cancellations in the calculation of the electron and neutron
EDMs in the small tan$\beta$ case. It is worth to extend the analysis to the large tan$\beta$ case
and investigate its effects on rare B decays.

This research was supported in part by the National Natural Science Foundation of China and 
partly supported by Center of Chinese Advanced Science and Technology(CCAST).
 


\newpage
\begin{figure}[t]
\vspace{0cm}
\centerline{
\epsfig{figure=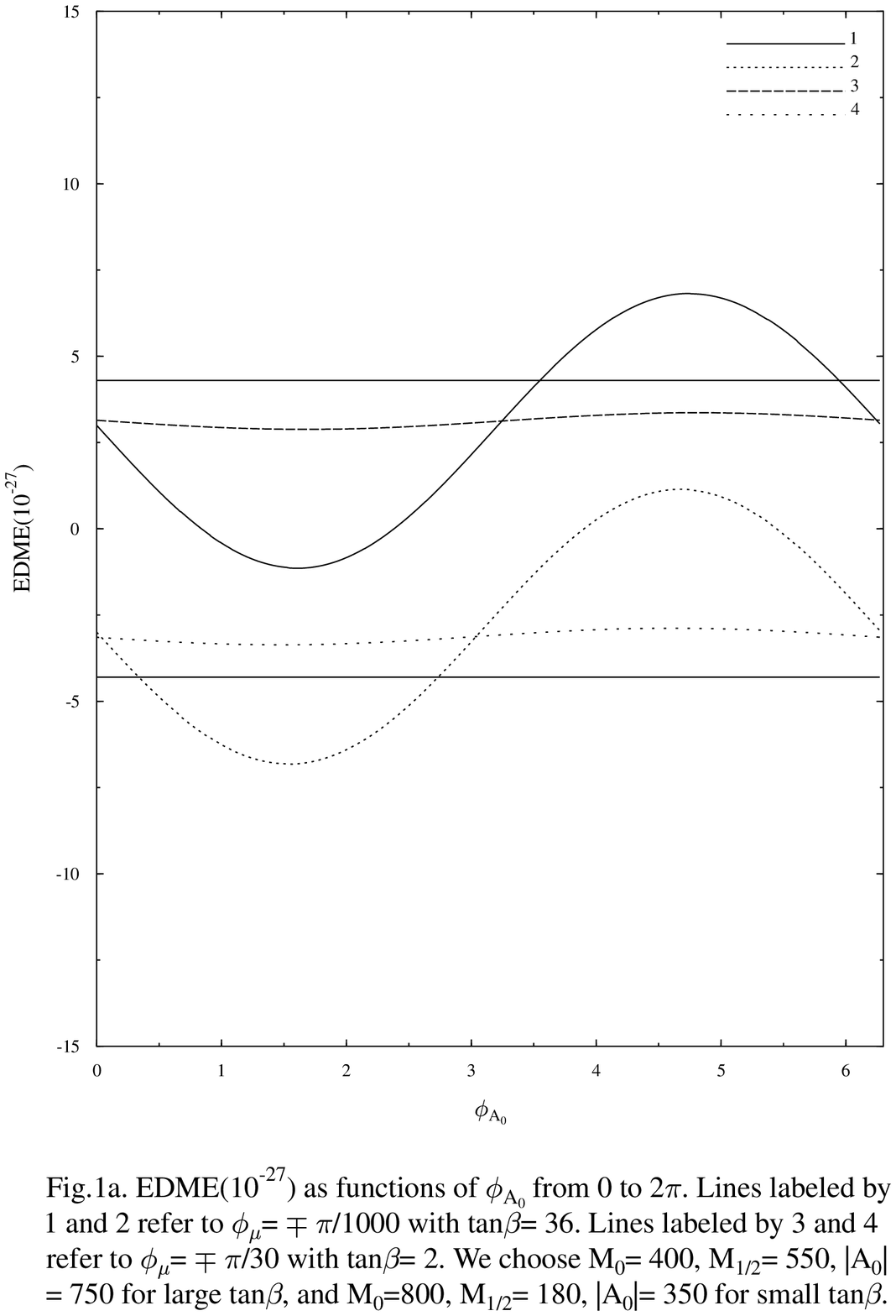,width=7.2 in}
}
\label{fig1a}
\end{figure}

\newpage
\begin{figure}[t]
\vspace{0cm}
\centerline{
\epsfig{figure=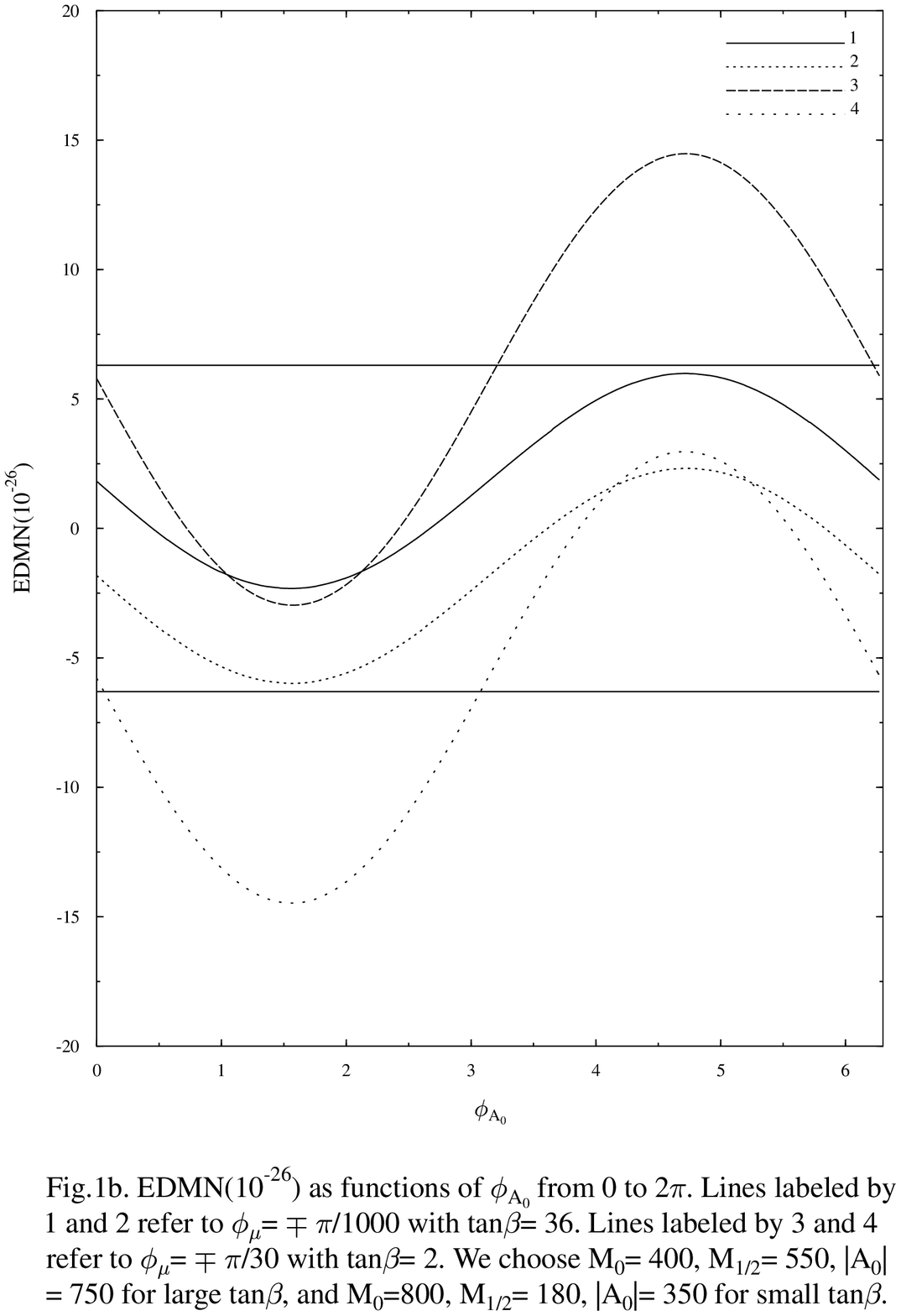,width=7.2 in}
}
\label{fig1b}
\end{figure}

\newpage
\begin{figure}[t]
\vspace{0cm}
\centerline{
\epsfig{figure=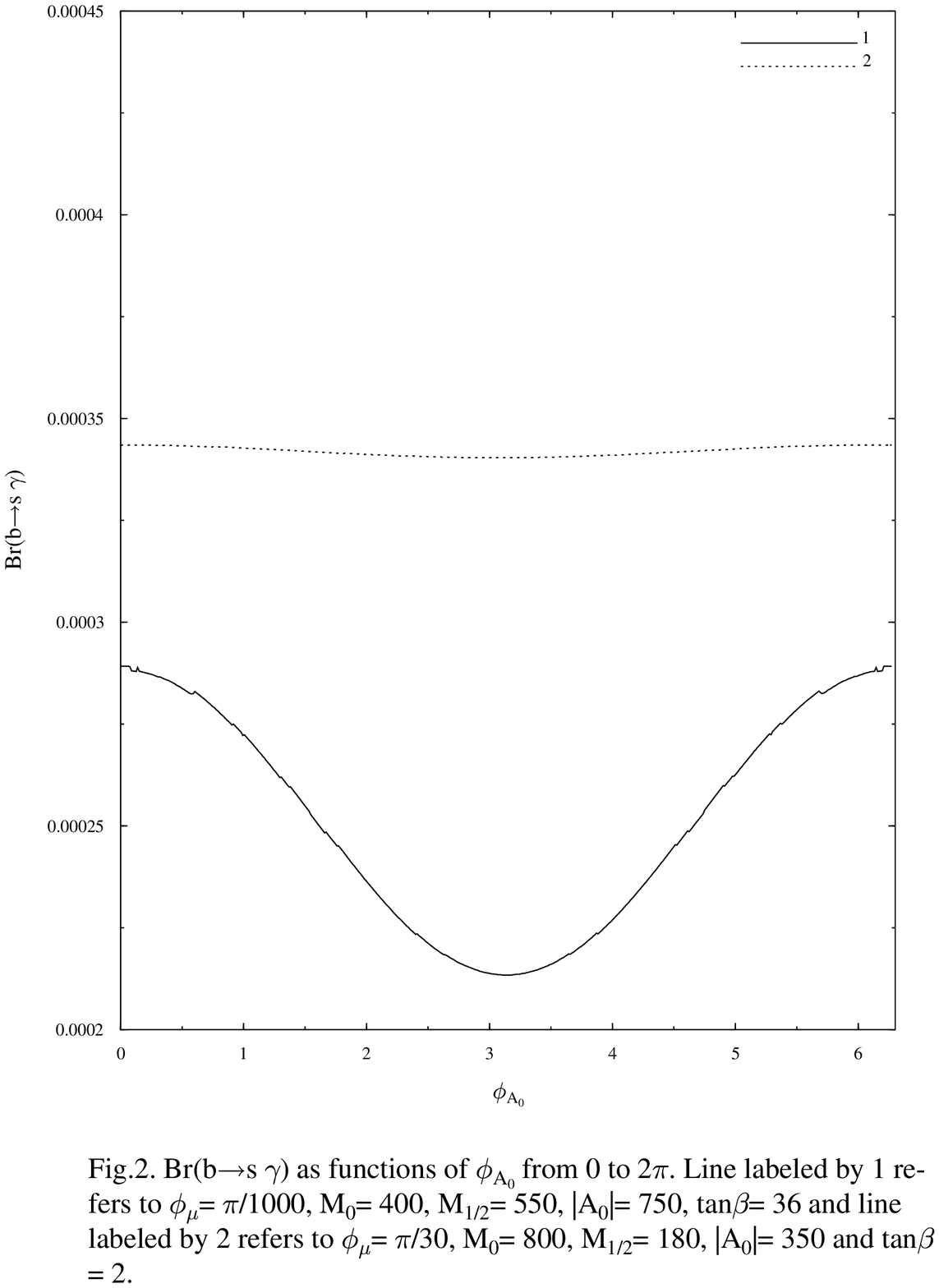,width=7.2 in}
}
\label{fig2}
\end{figure}

\newpage
\begin{figure}[t]
\vspace{0cm}
\centerline{
\epsfig{figure=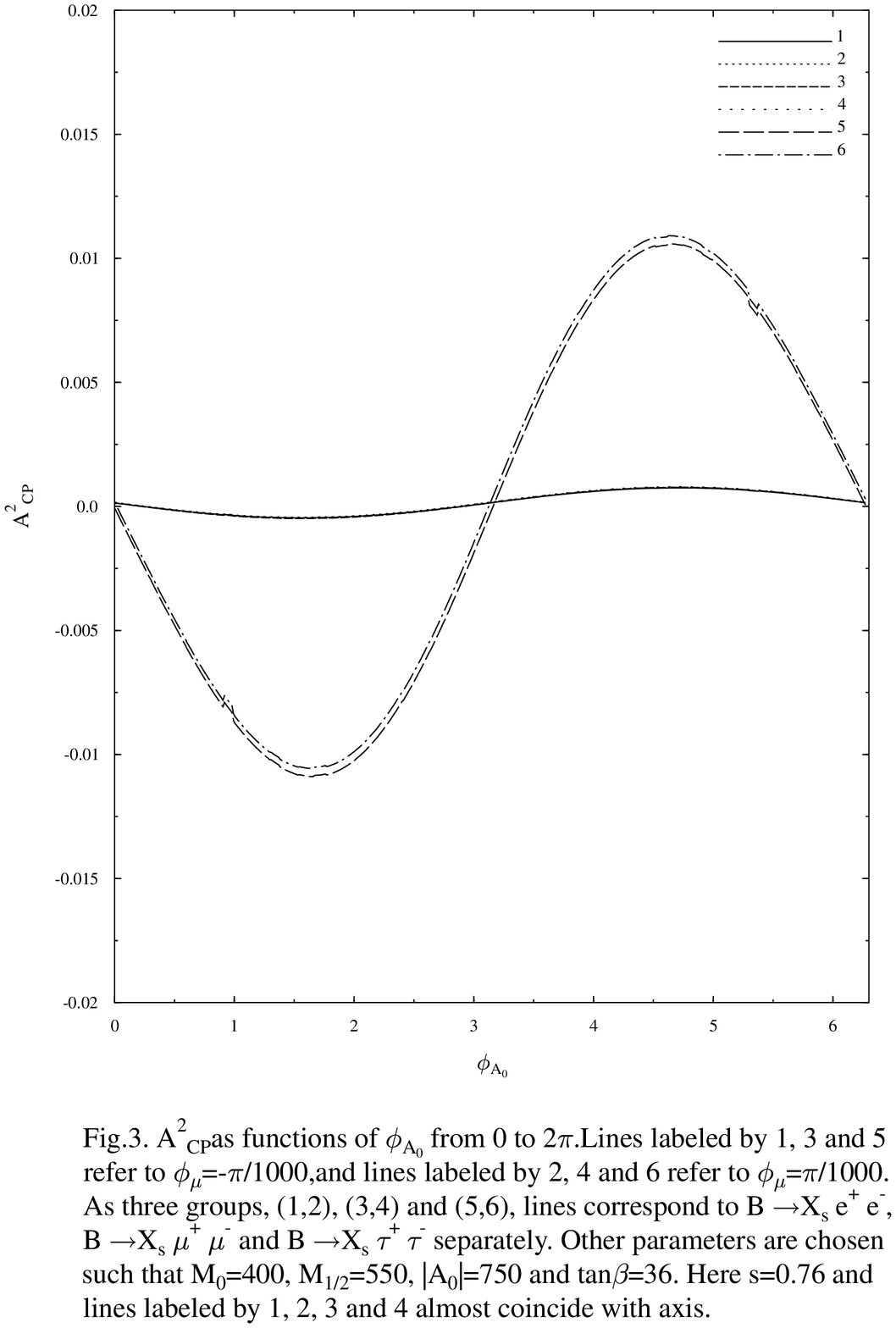,width=7.2 in}
}
\label{fig3}
\end{figure}

\newpage
\begin{figure}[t]
\vspace{0cm}
\centerline{
\epsfig{figure=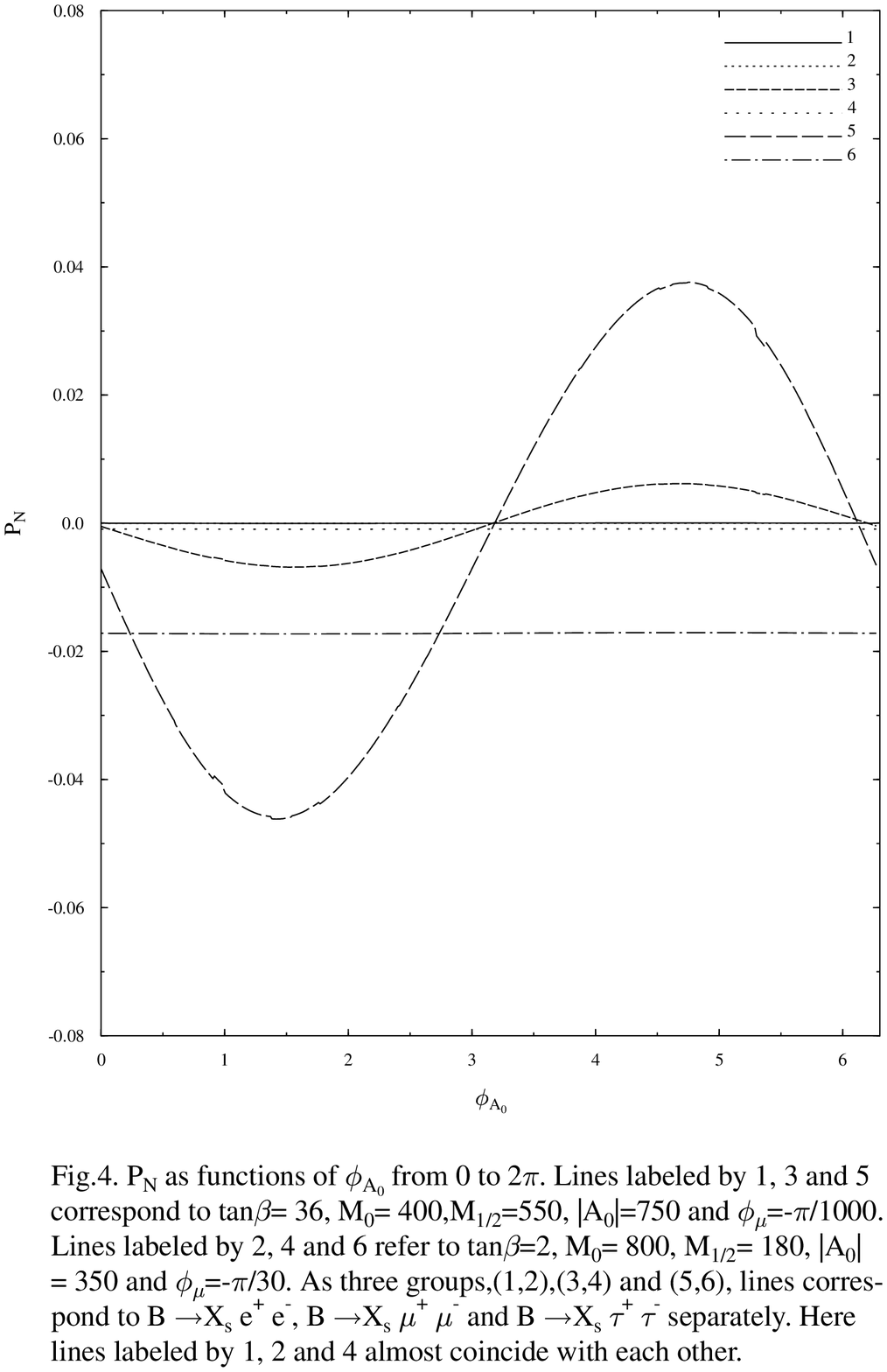,width=7.2 in}
}
\label{fig4}
\end{figure}


\begin{thebibliography}{99}
\bibitem{ckm}M. Kobayashi and T. Maskawa, Prog. Theor. Phys. {\bf 49} (1973) 652.
\bibitem{dgh} M.Dugan,B.Grinstein, and L.J.Hall,Nucl.Phys.B{\bf 255},
413 (1985); S.Dimopoulos and S.Thomas,Nucl.Phys. B{\bf 465},23(1996).
\bibitem{edm}J. Ellis, S. Ferrara and D. V. Nanopoulos, Phys. Lett. {\bf B114} (1982) 231;
for a review see S. M. Barr and W. J. Marciano, in CP Violation, edited by C. Jarlskog (World
Scientific, Singapore, 1989), p. 455; R. Barbieri, A. Romanino, and A. Strumia, Phys. Lett.
{\bf B369} (1996) 283.
\bibitem{nat}T. Falk, K.A. Olive, and M. Srednicki, Phys. Lett. {\bf B354} (1995) 99;
P. Nath, Phys. Rev. Lett. {\bf 66} (1991) 2565; Y. Kizukuri and N. Oshimo, Phys.
Rev. {\bf D45} (1992) 1806, {\bf D46} (1992) 3025; A. G. Cohen, D.B. Kaplan, A.E. Nelson,
Phys. Lett. {\bf B388} (1996) 588; A.G. Cohen et al., Phys. Rev. Lett. {\bf 78} (1997) 2300.
\bibitem{IN} T.Ibrahim and P.Nath, Phys.Rev. D{\bf 57},478(1998); (E) ibid, {\bf D58}, 019901
(1998); Phys. Rev. {\bf D58},111301(1998). 
\bibitem{bgk}M.Brhlik,G.J.Good,and G.L.Kane, Phys. Rev. {\bf D59},11504(1999); 
M.Brhlik, L. Everett, G.L.Kane, and J. Lykken, hep-ph/9908326.
\bibitem{FO}T. Falk and K.A. Olive, Phys. Lett. {\bf B375} (1996) 196, ibid, {\bf 439} (1998)
71.
\bibitem{goto}T. Goto et al., hep-ph/9812369.
\bibitem{aad}E. Accomando, R. Arnowitt and B. Dutta, hep-ph/9907446.
\bibitem{bz}S.M.Barr and A.Zee, Phys.Rev.Lett.{\bf 65}, 21(1990) 
\bibitem{ckp}D. Chang, W.-Y. Keung, A. Pilaftsis, Phys. Rev. Lett. {\bf 82} (1999) 900.
\bibitem{gk} Y.G. Kim, P. Ko and J.S. Lee, Nucl. Phys. {\bf
B544} (1999) 64; S. Back and P. Ko, hep-ph/9904283.
\bibitem{rge}K. Inoue et al., Prog. Theor. Phys. 68 (1982) 927; A. Bouquet, J. Kaplan and C.A. 
Savoy, Nucl. Phys. B 262 (1985) 299; V. Barger, M. Berger and P. Ohmann, Phys. Rev. {\bf D49} (1994) 4908.
\bibitem{data}E. Commins et al., Phys. Rev. {\bf A 50} (1994) 2960; K. Abdullah et al., Phys. Rev. Lett.
{\bf 65} (1990) 2347; P.G. Harris et al., Phys. Rev. Lett. {\bf 82} (1999) 904.
\bibitem{bg}R. Barbieri and G.F. Giudice Phys Lett B 309(1993)86;
M.A. Diaz, Phys. Lett. B 322 (1994) 591;
T. Goto and Y. Okada, Prog. of Theor. Phys. 94 (1995) 407;
P. Nath and R. Amorvitt, CERN-TH 7214/94, NUB-TH 3093/94, CTP-TAMU-32/94;
R. Garisto and J.N. Ng Phys. Lett. B 315 (1993) 372;
J.L. Hewett, SLAC-PUB-6521 May 1994 T/E;
F.M. Borzumati, M. Olechowski and S. Pokorski, CERN-TH 7515/94, TUM-73-83/94;
J.L. Lopez, D.V. Nanopoulos, X. Wang and A. Zichichi, Phys. Rev. D 51(1995) 147.
\bibitem{dhh} Y.B.Dai, C.S.Huang and H.W.Huang, Phys.Lett. B{\bf 390}
 257(1997).
\bibitem{hy}C.S. Huang and Q.S. Yan, Phys. Lett. {\bf B442} (1998) 209; C.S. Huang, Liao Wei, and Q.S. Yan, Phys. Rev.
{\bf D59} (1999) 011701.
\bibitem{gsw}B. Grinstein, M.J. Savage and M.B. Wise, Nucl. Phys. {\bf B319} (1989) 271.
\bibitem{h} C.S. Huang,  Commun. Theor. Phys.  {\bf  2}  (1983) 1265.
\bibitem{bb}S. Bertolini, F. borzumati, A. Masieso and G. Ridolfi, Nucl. Phys. B 353 (1991) 591;
P. Cho, M. Misiak and D. Wlyer, Phys. Rev. D 54 (1996) 3329.
\bibitem{Lee}T.D. Lee and C.S. Wu, Annu. Rev. Nucl. Sci. {\bf 16} (1966) 471.
\bibitem{KSAH} A. Ali and G. Hiller, Eur.Phys.J. {\bf C8}(1999) 619-629;
 F. Kruger and L.M. Sehgal, Phys.Rev.{\bf D55}(1997) 2799.
\bibitem{11} S.Rai Choudhury et al., hep-ph/9902355, where $P_N$  
have been given,but they gave only two terms in the numerator of $P_N$.
\bibitem{bs}Belle Progress Report, Belle Collaboration, KEK- PROGRESS-REPORT-97-1 (1997);
Status of the BaBar Detector, BaBar Collaboration, SLAC-PUB-7951, presented at 29th International
Conference on High Energy Physics, Vancouver, Canada, 1998.
\end{thebibliography}
\end{document}